\documentclass[fleqn,twocolumn,10pt]{wlscirep}
\usepackage[utf8]{inputenc}
\usepackage[T1]{fontenc}
\title{Characterization of quantum and classical correlations in the Earth's curved space-time}

\author[1]{Tonghua Liu}
\author[1,*]{Shuo Cao}
\author[2,$\dagger$]{Shumin Wu}

\affil{Department of Astronomy, Beijing Normal University, Beijing 100875, China\\
$^2$ Department of Physics, and Collaborative Innovation Center for Quantum Effects \\
and Applications, Hunan Normal University, Changsha, Hunan 410081,
China}

\affil[*]{caoshuo@bnu.edu.cn}
\affil[$\dagger$]{849588127@qq.com}

\begin{abstract}

The preparation of quantum system and the execution of quantum information tasks between distant users are always affected by gravitational and relativistic effects. In this work, we quantitatively analyze how
the curved space-time background of the Earth affect the classical
and quantum correlations of photon pairs, which are initially
prepared in a two-mode squeezed state. More specifically,
considering the rotation of the Earth, the space-time around the
Earth is described by Kerr metric. Our results show that these state
correlations, which initially increase for a specific range of
satellite's orbital altitude, will gradually approach a finite value
with increasing height of satellite's orbit (when the special
relativistic effect is involved). More importantly, our analysis
demonstrates that, the changes of correlations generated by the
total gravitational frequency shift could reach the level of
$<0.5\%$, within the satellite height at geostationary Earth orbits.

\end{abstract}
\begin{document}

\flushbottom
\maketitle
%
%
\thispagestyle{empty}

\section*{Introduction}

As one of the most important developments in modern physics, Quantum
entanglement -- a central characteristics of quantum correlations --
has arouse widespread attention in most recent years
\cite{entanglement,entanglement2}. However, one interesting question
which still calls for consideration is, can the separable state
(i.e., not entangled) only be determined as a classically correlated
state without quantum correlations? Following this direction,
several recent studies have focused on such issue and revealed some
signatures of quantumness in separable states, which indicates its
tremendous potential in future quantum technology. Note that one of
such signatures is the so-called quantum discord, which captures
general quantum correlations even in the absence of entanglement in
a quantum state. Following the methodology firstly proposed
\cite{discord2,discord1,discord3} and identified as a resource for
computation \cite{discord4}, quantum discord originates from the
discrepancy between two classically equivalent definitions of mutual
information, which can be derived from a measure of the total
correlations in a quantum state. Although the nature of quantum
discord is still unknown, it is rewarding to investigate its
important role played in the realization of some quantum information
tasks, especially in the absence of quantum entanglement. As was
found in \cite{noentang1,noentang3}, certain quantum information
processing tasks can also be done efficiently, even without the
participation of quantum entanglement.

In realistic situation, the preparation of quantum system and the
procession of quantum information tasks are always accompanied with
gravitational and relativistic effects. Considering the fact that
the previous works have paid more attention to quantum discord
without gravitational or relativistic effects, its behaviors in a
relativistic setting or curved space-time background is still an
uncharted territory. Fortunately, the quantum field theory of curved
space-time provides a theoretical framework to carry out the above
analysis \cite{DEBA,DEBT}, which enables one to incorporate
relativistic effects into quantum experiments. Nevertheless, when
studying quantum resource in relativistic setting, the effects of
gravity and motion -- especially on the quantum properties and their
applications -- have always been ignored, which fails to overcome
the inherent inconsistency between quantum physics and relativity.
Such gap has recently been bridged by quantum field theory in curved
space-time, which was employed as the core framework to compute the
ultimate bounds on ultra-precise measurements of relativistic
parameters \cite{DEBA,DEBT}. Further progress in this direction has
been achieved by \cite{Darra,Sugumi} in two papers discussing
quantum discord in de Sitter space and between relatively
accelerated observers. Meanwhile, understanding the influence of
gravitational effects on quantum discord also has practical and
fundamental significance in realistic world, especially when the
parties involved are located at large distances in the curved
space-time \cite{DSSC}. For instance, it was found in \cite{SPK}
that there would be inevitable losses of quantum resources in the
estimation of the Schwarzschild radius. Furthermore, a quantitative
investigation of the dynamics of satellite-based quantum steering
and coherence has been presented in \cite{TQJJH,Laserliu}, given the
curved background space-time of the Earth. Nowadays, it is possible
to explore those quantum correlations (with quantum discord acting
as resources in quantum protocol) and the correlations in
relativistic quantum systems (which is related to the practical
implementation of many quantum information schemes, i.e., quantum
key distribution through satellite nodes \cite{nature} and other
quantum information tasks \cite{MJAG,HW,LV,WH,satellite1,satellite2}).
Therefore, it is practical and fundamental importance to study the influence of gravitational effects on the quantum resources when the parties involved  are located at long distances in the curved space-time.

In this work, focusing on the classical and quantum correlations and
its behaviors under gravitational effect of the Earth, we will
quantitatively analyze how the curved space-time background of the
Earth influence these correlations. The correlated photon pairs are
initially prepared in a two-mode squeezed state, one of which is
then assumed to stay at Earth's surface with the other propagating
to the satellite. Note that the photons' wave-packet will be
deformed by the curved background space-time of the Earth in the
propagating process, while a lossy quantum channel can be used to
model these deformed effects on the quantum state of photons
\cite{MANI}. Specially, we will quantitatively calculate the losses
of classical and quantum correlations, and furthermore discuss their
behaviors in the curved space-time of the Earth. The advantage of this
work is that the Earth’s gravitational field is described as lossy channel rather than global free models, because the latter suffers from the single-mode approximate problem and physically unfeasible detection in the full space-time. Meanwhile, according to the equivalence principle, the effects of acceleration are equivalence with the effects of
gravity, our works could be in principle apply to all types
of correlations affected by the acceleration field.

This paper is
organized as follows. Firstly, we describe the quantum field theory
of a massless uncharged bosonic field propagating from the Earth to
a satellite. Secondly, we briefly introduce the definition of the
measurements of mutual information, classical correlation, and
quantum correlation (or the quantum discord) for a bipartite
Gaussian state. Thirdly, we show a scheme to test large distance
quantum discord between the Earth and satellites, based on which the
behaviors of three types of correlations will be studied in the
curved space-time. Throughout the paper we employ natural units $G =
c =\hbar= 1$.

\section*{Results}

\subsection*{Light wave-packets propagating in the Earth's space-time}

In this section, we will give a brief introduction about the
propagation of photons from the Earth to satellites, under the
influence of the Earth's gravitational field. Considering the
rotation of the Earth, the space-time considered in this analysis
can be approximately described by the Kerr metric \cite{Visser}. For
the sake of simplicity, our work will be constrained to the
equatorial plane and the Kerr line element in Boyer-Lindquist
coordinates $(t,r,\phi)$ reduces to \cite{Visser}
\begin{align}\label{metric}
ds^2=&\, -\Big(1-\frac{2M}{r} \Big)dt^2+\frac{1}{\Delta}dr^2 \nonumber \\
&\,+\Big(r^2+a^2+\frac{2Ma^2}{r}\Big) d\phi^2 - \frac{4Ma}{r} dt \, d\phi, \\
\Delta=&\,1-\frac{2M}{r}+\frac{a^2}{r^2},
\end{align}
where $M$ and $r$ respectively denote the mass and radius of the
rotating planet. Based on the angular momentum $J$, the Kerr
parameter (i.e., normalized angular momentum) can be expressed as
$a=\frac{J}{M}$.

In order to clearly describe the propagation of wave-packets from a
source on Earth to a receiver satellite situated at a fixed
distance, two observers called Alice and Bob are prepared as the
reference frames, respectively. More specifically, focusing on a
photon sent from Alice at the time of $\tau_A$, it will arrive at
Bob at the time of $\tau_B=\Delta\tau+\sqrt{f(r_B)/f(r_A)}\tau_A$,
where $f(r)$ is the gravitational frequency shifting factor and the
$\Delta\tau$ represents the propagation time of the light from Alice
to Bob. As is well known, photons can be modeled by the wave packet
of excitations of massless bosonic field, with a distribution of
$F^{(K)}_{\Omega_{K,0}}$ ($\Omega_{K}$ is the mode frequency peaked
at $\Omega_{K,0}$)\cite{ULMQ,TGDT}, where $K=A, B$ denote the modes
in Alice's or Bob's reference frames, respectively. For an observer
infinitely far from Alice or Bob, the annihilation operator for the
photon takes the form
\begin{equation}
\hat{a}_{\Omega_{K,0}}(t_K)=\int_0^{+\infty}d\Omega_K e^{-i\Omega_K
t_K}F^{(K)}_{\Omega_{K,0}}(\Omega_K)\hat{a}_{\Omega_K}, \label{wave}
\end{equation}
with the frequency distribution $F^{(K)}(\Omega)$, which is
naturally applied to model a photon, a wave packet of the
electromagnetic field localized and propagating in space-time. Note
that the creation $\hat{a}^{\dagger}_{\Omega_{K,0}}$ and
annihilation $\hat{a}_{\Omega_{K,0}}$ operators satisfy the
canonical equal time bosonic commutation relations
($[\hat{a}_{\Omega_{K,0}}(t),\hat{a}^{\dagger}_{\Omega_{K,0}}(t)]=1$)
when the frequency distribution $F^{(K)}(\Omega)$ is normalized
(i.e., $\int_{\Omega>0}|F^{(K)}(\Omega)|^2=1$).

\begin{figure}[tbp]
\centering
\includegraphics[height=4.6in, width=3.6in]{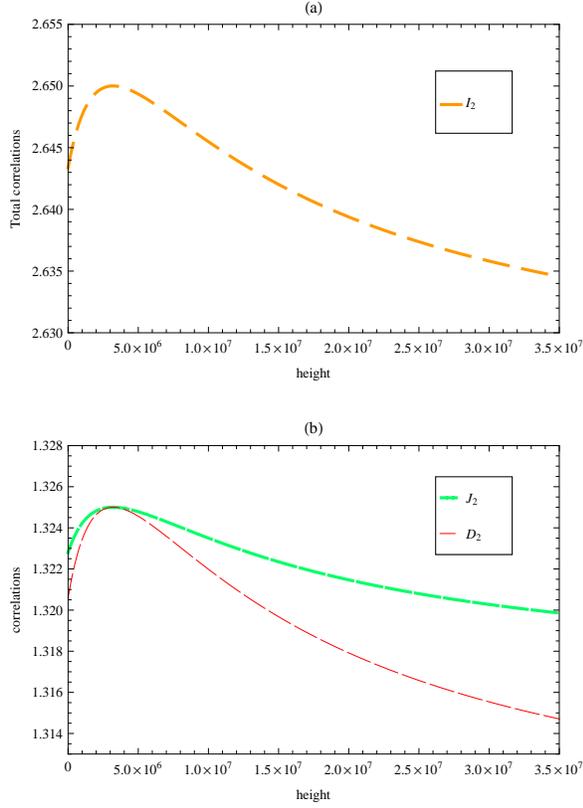}
\caption{(Color online) Three types of correlations ${I}_2$
(Yellow), ${J}_2$ (Green) and ${D}_2$ (Red) as
functions of increasing orbit height $h$. The Gaussian bandwidth,
squeezing parameter and frequency of mode $b_2$ are fixed at
$\sigma=1$, $s=1$ and $\Omega_B=1$, respectively.}
\end{figure}

\begin{figure}[tbp]
\centering \centerline{\includegraphics[width=8.2cm]{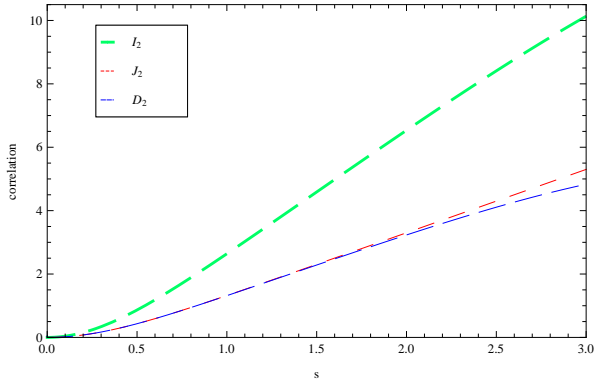}}
\caption{(Color online) Three types of correlations ${I}_2$
(Green), ${J}_2$ (Red) and ${D}_2$ (Blue) as
functions of the squeezing parameter $s$. The orbit height of the
satellite, frequency of mode $b_2$ and the Gaussian bandwidth are
fixed at $h=2\times10^4$ km, $\Omega_B=1$ and $\sigma=1$,
respectively. }\label{f1}
\end{figure}

Now let us consider a realistic situation, in which Alice (located
on the surface of the Earth, $r_A=r_E$) prepared and sent a wave
packet $F^{(A)}_{\Omega_{A,0}}$ to Bob (located on the satellite),
then the latter received this wave packet $F^{(B)}_{\Omega_{B,0}}$
with an altitude of $r_B$. Considering the Earth's gravitational
field, the wave packet received by Bob should be modified, following
the relation between the two wave packet of frequency distributions
\cite{DEBT,DEBA}. Moreover, the gravity of the Earth will also
change the mode frequency $\Omega_K$ with the shift of
$\Omega_A=\sqrt{f(r_A)/f(r_B)}\Omega_B$, where $f(r_A)/f(r_B)$ is
the total gravitational frequency shifting function (see more
details in the text). Therefore, the total modified contribution
induced by the gravitational effects of the Earth can be written as
\begin{eqnarray}
F^{(B)}_{\Omega_{B,0}}(\Omega_B)=\sqrt[4]{\frac{f(r_B)}{f(r_A)}}F^{(A)}_{\Omega_{A,0}}\left(\sqrt{\frac{f(r_B)}{f(r_A)}}\Omega_B\right).\label{wave:packet:relation}
\label{fab}
\end{eqnarray}
One may clearly see that the effect induced by the curved space-time
of the Earth cannot be simply corrected by a linear shift of
frequency, which indicates its great difficulty in compensating such
transformation induced by the curvature in realistic
implementations.

Fortunately, following the relation between such nonlinear
gravitational effect and the fidelity of the quantum channel
\cite{DEBT,DEBA}, it is possible to decompose the mode
$\bar{a}^{\prime}$ received by Bob in terms of the mode $a^{\prime}$
prepared by Alice and an orthogonal mode $\hat{a}_{\bot}^{\prime}$
(i.e., $[\hat{a}^{\prime},\hat{a}_{\bot}^{\prime\dagger}]=0$)
\cite{PPRW}
\begin{eqnarray}
\bar{a}^{\prime}=\Theta
\hat{a}^{\prime}+\sqrt{1-\Theta^2}\hat{a}_{\bot}^{\prime}.\label{mode:decomposition}
\end{eqnarray}
Here $\Theta$ is the wave packet overlap between the distributions
$F^{(B)}_{\Omega_{B,0}}(\Omega_B)$ and
$F^{(A)}_{\Omega_{A,0}}(\Omega_B)$, which takes the form of
\begin{eqnarray}
\Theta:=\int_0^{+\infty}d\Omega_B\,F^{(B)\star}_{\Omega_{B,0}}(\Omega_B)F^{(A)}_{\Omega_{A,0}}(\Omega_B).\label{single:photon:fidelity}
\end{eqnarray}
It is easy to see that $\Theta=1$ corresponds to a perfect channel,
while $\Theta<1$ represents a noisy channel under the influence of
the Earth's curved space-time, the quality of which can be
quantified by employing the fidelity of ${F}=|\Theta|^2$. In
order to better characterize the frequency distribution of the
source, a real normalized Gaussian wave packet is applied to Alice's
mode
\begin{eqnarray}
F_{\Omega_0}(\Omega)=\frac{1}{\sqrt[4]{2\pi\sigma^2}}e^{-\frac{(\Omega-\Omega_0)^2}{4\sigma^2}}\label{Bobpacket},
\end{eqnarray}
with the wave packet width $\sigma$. We remark here that in the
expression of the overlap parameter $\Theta$
(Eq.~\eqref{single:photon:fidelity}), the domain of integration will
be extended to all real axis and the integral should be performed
over strictly positive frequencies. This is justified by the fact
that the peak frequency is typically much larger than the spreading
of the wave packet (i.e.,$\Omega_0\gg \sigma$). Based on the
combination of Eqs. \eqref{fab} and \eqref{Bobpacket}, one could
obtain
\begin{eqnarray} \label{theta}
\Theta=\sqrt{\frac{2}{1+(1+\delta)^2}}\frac{1}{1+\delta}e^{-\frac{\delta^2\Omega_{B,0}^2}{4(1+(1+\delta)^2)\sigma^2}}\label{final:result},
\end{eqnarray}
where the new parameter $\delta$ is introduced to quantify the
shifting effect:
\begin{equation}\label{aw}
\delta=\sqrt[4]{\frac{f(r_A)}{f(r_B)}}-1=\sqrt{\frac{\Omega_B}{\Omega_A}}-1.
\end{equation}
In the equatorial plane of the Earth described by Kerr metric, the
expression of parameter $\frac{\Omega_B}{\Omega_A}$ is rewritten as
\cite{kerr}
\begin{equation}
\frac{\Omega_B}{\Omega_A}=\frac{1+\epsilon
\frac{a}{r_B}\sqrt{\frac{M}{r_B}}}{C\sqrt{1-3\frac{M}{r_B}+
2\epsilon\frac{a}{r_B}\sqrt{\frac{M}{r_B}}}}.
\end{equation}
Here the normalization constant takes the form of
$C=[1-\frac{2M}{r_A}(1+2a
{\omega})+\big(r^2_A+a^2-\frac{2Ma^2}{r_A}\big){\omega}^2]^{-\frac{1}{2}}$
(with the Earth's equatorial angular velocity $\omega$), and
$\epsilon=\pm1$ stands for the direction of orbits (i.e., the
satellite co-rotates with the Earth when $\epsilon=+1$).

In order to obtain the explicit expression of the frequency shift
for the photon exchanged between Alice and Bob, we try to expand the
Eq. (\ref{aw}) by keeping the first order of the perturbation term
$(r_A \omega)^2$. Note that the perturbative result does not depend
on the state of the Earth and the satellite (i.e., whether they are
co-rotating) and the shift parameter is expressed as
\begin{eqnarray}\label{bw}
\nonumber\delta&=&\delta_{Sch}+\delta_{rot}+\delta_h\\
\nonumber&=&\frac{1}{4}\frac{r_S}{r_A}\big(\frac{r_A-2h}{r_A+h} \big)-\frac{(r_A\omega)^2}{2}-\frac{(r_A\omega)^2}{4}\big(\frac{3}{4}\frac{r_S}{r_A}-\frac{2r_Sa}{\omega r_A^3}\big),
\end{eqnarray}
where $\delta_{Sch}$, $\delta_{rot}$, and $\delta_h$ respectively
denote the first order Schwarzschild term, the rotation term, and
the higher-order correction term. Note that $r_S=2M$ is the
Schwarzschild radius of the Earth, while the parameter $h=r_B-r_A$
quantifies the height difference between Bob and Alice. It is
necessary to mention that due to the effects induced by the rotation
of the Earth, the overlap parameter $\Theta$ is no longer equal to
one when $h=\frac{r_A}{2}$, which is quite different from the
Schwarzschild case. However, when the satellite moves at the height
of $h\simeq\frac{r_A}{2}$, the gravitational effect of the Earth and
the special relativistic effect (i.e., doppler effect) will
compensate each other ($\Theta=1$), due to the motion of the
satellite. Therefore, the photons received by Bob at this height
will not experience any frequency shift, which implies that Bob's
clock rate will become equal to that of Alice.

\subsection*{The influence of Earth's curved space-time on three types of correlations}

In this section, in the framework of photon exchange between the
ground and satellites, we propose a scheme to test the classical and
quantum correlations at large distance, and furthermore quantify the
effect of the Earth's curved space-time on such correlations. In the
framework of a pair of entangled photons initially prepared in a
two-mode squeezed state (with the modes of $b_1$ and $b_2$ at the
ground station), we firstly send mode a photon with mode $b_1$ to
Alice, with the other photon with mode $b_2$ propagating from the
Earth to the satellite and then received by Bob. Due to the curved
background space-time of the Earth, the wave packet of photons will
be deformed. More specifically, the total correlation, classical
correlation, and quantum correlation (quantum discord) of this
system will be investigated in the curved space-time.

Considering the fact that Alice and Bob receive the two modes ($b_1$
and $b_2$) at different satellite orbits, the effects induced by the
curved space-time should be included in the total process. As was
discussed in \cite{DEBT,DEBA}, such space-time effects on the
two-mode squeezed state can be modeled by a beam splitter with
orthogonal modes $b_{1\bot}$ and $b_{2\bot}$. The covariance matrix
of the initial two-mode squeezed state is given by
\begin{equation}\label{initialcov}
\Sigma^{b_1b_2b_{1\bot}b_{2\bot}}_0=\left(
\begin{array}{cc} \tilde\sigma(s) &0 \\ 0  &  I_4
\end{array}\right),
\end{equation}
Here ${I}_4$ denotes the $4\times4$ identity matrix and
$\tilde\sigma{(s)}$ is the covariance matrix of the two-mode
squeezed state
\begin{equation}
\tilde\sigma(s)=\left(
\begin{array}{cc} \cosh{(2s)}  {I}_2&\sinh{(2s)}\sigma_z \\ \sinh(2s)\sigma_z &\cosh{(2s)}  {I} _2
\end{array}\right),
\end{equation}
with the Pauli matrix $\sigma_z$ and squeezing parameter $s$.
Following the recent analysis of \cite{DEBT,DEBA}, one can use lossy
channels to model the effects of the curved space-time on Alice's
mode $b_1$ and Bob's mode $b_2$, which are described by the
transformation. In our scheme, we only consider Bob's mode sent to
the satellite and Alice's mode sent to the ground, which means that
Alice suffers a perfect channel ($\Theta_1=1$). Therefore, one may
naturally obtain
\begin{eqnarray}
\bar{b}_1&=&\,b_1\\ \nonumber
\bar{b}_2&=&\Theta_2\,b_2+\sqrt{{1-\Theta_2^2}}b_{2\bot},
\end{eqnarray}
and a mixing beam splitting of modes $b_1(b_2)$ and
$b_{1\bot}(b_{2\bot})$ will represent this process. For the entire
state, the symplectic transformation can be encoded into the
Bogoloiubov transformation
\begin{equation}
S=\left(
\begin{array}{cccc}
  {I}_2 &0&  0&0  \\ 0&\Theta_2  {I} _2 &0&\sqrt{
1-\Theta_2^2}  {I} _2\\  0 &0&  -{I}_2&0  \\ 0&\sqrt{
1-\Theta_2^2}  {I} _2 &0&-\Theta_2  {I} _2
\end{array}\right),\nonumber
\end{equation}
based on which the final state $\Sigma^{b_1b_2b_{1\bot}b_{2\bot}}$
after transformation is
$\Sigma^{b_1b_2b_{1\bot}b_{2\bot}}=S\,\Sigma_0^{b_1b_2b_{1\bot}b_{2\bot}}\,S^{T}$.
Then we trace over the orthogonal modes ($b_{1\bot},b_{2\bot}$) and
obtain the covariance matrix $\Sigma^{b_1b_2}$ for the modes ($b_1$
and $b_2$) after propagation
\begin{equation}\Sigma^{b_1b_2}=\left(
\begin{array}{cc}\label{gst}
(1+2\sinh^2s ) {I}_2 &\sinh{(2s)}\,\Theta_2\,\sigma_z  \\ \sinh{(2s)}\,\Theta_2\,\sigma_z &(1+2\sinh^2s\,\Theta_2^2 )\, {I}_2
\end{array}\right).
\end{equation}

\begin{figure}[tbp]
\centerline{\includegraphics[height=4.6in, width=3.6in]{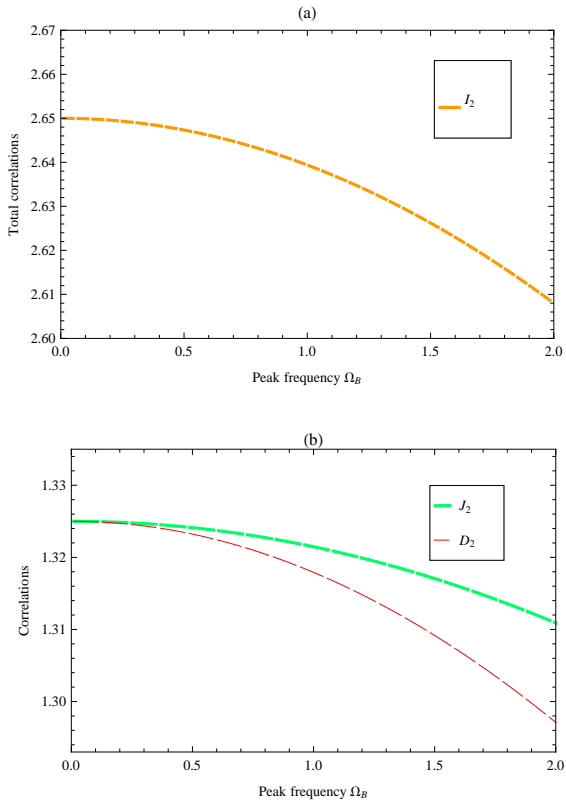}}
\caption{ (Color online). Three types of correlations
${I}_2$ (Yellow), ${J}_2$ (Green) and
${D}_2$ (Red) in terms of the peak frequency of mode $b_2$.
The other parameters are fixed at $s=1$, $\sigma=1$ and
$h=2\times10^4$ km. }\label{f1}
\end{figure}

\begin{figure}[tbp]
\centering
\includegraphics[height=2.2in, width=3.2in]{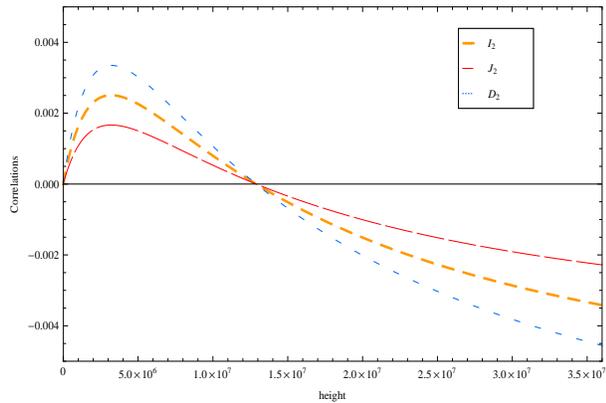}
\caption{(Color online) The change rate of three types of
correlation $\mu_C$ in term of the satellite's orbital height. The
other state parameters are fixed at $s=1$, $\sigma=1$ and
$\Omega_B=1$. }
\end{figure}

Now the final form of the two-mode squeezed state, which suffers
from the influence of the Earth's gravity can be characterized by
Eq. (\ref{gst}). Let's remark here that, on the one hand, a lossy
quantum channel determined by the wave packet overlap parameter
$\Theta$ (which contains parameters $\delta$, $\sigma$ and
$\Omega_{B,0}$), could fully describe the effect of the Earth's
curved space-time on the quantum state. On the other hand, one could
straightforwardly obtain $\delta\sim 2.5\times10^{-10}$, considering
the Schwarzschild radius of the Earth ($r_S=9$ mm) and a typical
case that satellites always stay within geosynchronous satellite
orbit. Moreover, a typical parametric down converter crystal (PDC)
source with a wavelength of 598 nm (corresponding to the peak
frequency $\Omega_{B,0}= 500$ THz) and Gaussian bandwidth
$\sigma=1$MHz are also considered in our case \cite{NCMS, DNMP},
from which one could derive
$\delta\ll(\frac{\Omega_{B,0}}{\sigma})^2\ll1$. Therefore, for
simplicity and clarity, we expand the wave packet overlap $\Theta$
by keeping the second-order terms of the parameter $\delta$, i.e.,
$\Theta \sim1-\frac{\delta^2\Omega_{B,0}^2}{8\sigma^2}$. In order to
ensure the validity of perturbation expansion for the Matrix
element, we estimate the value of
$\frac{\delta^2\Omega^2_{B,0}}{8\sigma^2}\sim3.2\times10^{-8}$,
based on which one may argue that the perturbative expansion is
valid when the value of the squeezing parameter is $s\ll7.6$
(corresponding to $\sinh^2(s)\ll10^6$). Therefore, we can safely
carry out the following work with the squeezing parameter $s<3$. For
convenience, we will work with dimensionless quantities by rescaling
the peak frequency and the Gaussian bandwidth
\begin{equation}
\tilde{\Omega}\equiv\frac{\Omega}{\Omega_{B,0}}, \tilde{\sigma}\equiv\frac{\sigma}{\sigma_0},
\end{equation}
where $\Omega_{B,0}=500$ THz and $\sigma_0=1$ MHz. For simplicity,
the dimensionless parameters $\tilde{\Omega}$ and $\tilde{\sigma}$
are abbreviated as $\Omega$ and $\sigma$, respectively.

Now the effect of the Earth's curved space-time will quantified by
three types of correlations. According to
Eq.~(\ref{Total})-(\ref{quantum}), we could obtain the behaviors of the total correlation ${I}_2$, classical
correlation ${J}_2$ and quantum discord ${D}_2$
between the two modes of $b_1$ and $b_2$. The corresponding results
are explicitly shown in in Fig.~1, which illustrates the three types
of correlations as functions of increasing orbit height $h$. The
Gaussian bandwidth, squeezing parameter and frequency of mode $b_2$
are respectively fixed at $\sigma=1$, $s=1$ and $\Omega_B=1$. One
may clearly see that, compared with the classical correlation, the
quantum discord will be more easily effected by the Earth's curved
space-time. Such tendency has been firstly noted and extensively
discussed in the previous works \cite{QuantumGrav}. Moreover, the
three type of correlations between mode $b_1$ and $b_2$ have
exhibited the similar behavior, i.e., initially increase for a
specific range of height parameter $h\simeq\frac{r_A}{2}$ and then
gradually approach to a finite value with increasing $h$. One
possible explanation to such findings is based on the fact that
gravitational frequency shift and special relativistic effects play
different roles in reducing correlations. With the increase of
satellite height, the special relativistic effects becomes smaller
while the gravitational frequency shift can be cumulated. More
specifically, the photon's frequency received by the satellites at
the height of $h<\frac{r_A}{2}$ will experience blue-shift (with
increasing correlations), while the corresponding frequencies
received at the height $h>\frac{r_A}{2}$ will experience redshift
(with decreasing correlations). In fact, the peak value of all
correlations (i.e., the parameter $\delta=0$) has strongly suggested
a detectable frequency transformation from blue-shift to red-shift,
in the framework of three types of correlations between the photon
pairs \cite{kerr}. It should be pointed out that, when the two
parties are situated at the same height or in the flat space-time
($\delta\neq 0$), the total frequency shift generated by both
special and general relativistic effects should be taken into
account (Eq.~(\ref{aw}))\cite{kerr}. When the satellite moves at the
height $h=\frac{r_A}{2}$ accompanied with vanishing Schwarzschild
term ($\delta_{Sch}$), the photons received on satellites will
generate a tiny frequency shift, in the case of which the
lowest-order rotation term $\delta_{rot}$ and higher-order
correction term $\delta_{h}$ should be taken into consideration.

In order to better understand the relations between the three
different correlations and the initial squeezing parameter, we also
show the correlations of ${I}_2$, ${J}_2$ and
${D}_2$ changing with the squeezing parameter $s$, fixing
the orbit height $h$ at geostationary Earth orbits $3.6\times10^4$
km, the frequency of mode $b_2$ at 1, and the Gaussian bandwidth
$\sigma$ at 1. As can be clearly seen from Fig.~2, although all of
the three type of correlations will increase with the squeezing
parameter, the total correlation ${I}_2$ is much more
sensitive to the change of squeezing parameter, compared with the
classical correlation and the quantum discord. Similarly,
considering the degeneracy between the wave packet overlap parameter
$\Theta$ and the frequency parameter $\Omega_B$, it is also
necessary to investigate the behavior of three types of
correlations with $\Omega_B$. The three types of correlations as functions with the increasing frequency parameter $\Omega_B$ of mode $b_2$ are shown in Fig. 3. It clearly show that all types of correlations decrease with the
increasing frequency parameter. Moreover, quantum discord is easier to change with increasing squeezing parameter than classical correlation. We emphasize that compared with the
classical correlation, the quantum discord will slightly vary
with increasing frequency parameter, which strongly indicates the
possibility that one could choose appropriate parameters to realize
quantum communications from the Earth to the satellite.

Finally, with the aim of furthermore quantifying the influence of
the Earth's curved space-time, we will define an additional quantity
to describe the change rate of three types of correlations
\begin{equation}
\mu_C=\frac{C(\Sigma^{b_1b_2})-C_0(\Sigma^{b_1b_2})}{C_0(\Sigma^{b_1b_2})},
\end{equation}
where $C={I}_2, {J}_2,{D}_2$ and the
subscript $0$ denotes its corresponding value at the height of the
satellite ($h=0$). In Fig.~4, we plot the change rate of correlation
$\mu_C$ as a function of the height parameter $h$, with fixed
Gaussian bandwidth $\sigma=1$ and frequency parameter $\Omega_B=1$.
It is easy to perceive the effects of blue-shift in the range of
$h<\frac{r_A}{2}$ and red-shift at the height of $h>\frac{r_A}{2}$,
based on the behavior of the $\mu$ parameter in the framework of
three types of correlations. Therefore, the point of $h=2r_A$ (which
corresponds to $\delta=-1$) denotes a special height at which the
blue-shift and the red-shift effects on a photon's frequency are
offset. More specifically, the change of correlations generated by
gravitational frequency shift can be determined at the level of
$<0.5\%$, within the satellite height at geostationary Earth orbits.
Such findings could provide some interesting possibilities to reduce
the loss of three types of correlation (especially the quantum
discord), through the control of the orbital height of satellites.

\section*{Discussion}
In conclusion, we have studied the influence of curvature induced by the Earth on three types of correlations including total correlation (mutual information),
classical correlation, and quantum correlation (quantum discord) for
a two-mode Gaussian state, in which one of the modes is propagating
from the ground to satellites. Different from the Schwarzschild case
(no rotation) discussed in the previous works, the special
relativistic effects are also involved in our analysis, considering
the rotation of the Earth and the curved space-time described by
Kerr metric. Our results strongly indicate that all of the three
types of quantum correlations increase with the compression
parameters, while the mutual information is more sensitive to the
squeezing parameter $s$, compared with the classical correlation and
quantum discord. Meanwhile, focusing on the relations between the
correlations and frequency, the quantum discord will significantly
vary with increasing frequency parameter $\Omega_B$, which is quite
different from the behavior of the classical correlation. Compared with squeezing parameter, the frequency parameter $\Omega_B$ effecting on three types of correlations has inverse influence, which means that a lower peak frequency parameter will lead to less loss for all of correlations. However, compared with other input parameters, squeezing parameter are dominant, i.e., bigger means better.
Finally, in the framework of a quantity describing the change rate of three
types of correlations, we detect a special height ($h=2r_A$) at
which the blue-shift and the red-shift effects on a photon's
frequency are offset.  More importantly, the change of correlations
generated by gravitational frequency shift is determined at the
level of $<0.5\%$, within the satellite height at geostationary
Earth orbits.

It should be emphasized that, with the rapid developments in both
quantum technology and quantum communication, it is possible to
implement quantum task between the ground and satellites with the
three types of correlation studied in this work. More interestingly,
considering the fact that realistic quantum systems always exhibit
gravitational and relativistic features, our analysis in this paper
can be extended to the investigation of the dynamics of all types of
correlation under the influence of acceleration. Such conclusion is
supported by the equivalence principle in General Relativity, which
states that the effects of gravity are exactly equivalent to the
effects of acceleration. Moreover, realizations of quantum information and communication task must involve multi-particle systems in the future, if we assume the initial states is multi-particle quantum state, and then calculate covariance matrix of the initial state after the propagation, the final state can be obtained. In the above process, one ends up dealing with Gaussian states and transformations. Therefore, our works can be extended to multi-particle quantum state.

\section*{Methods}

For a general two-mode Gaussian state ($\rho_{AB}$) composed of two
subsystems ($A$ and $B$) \cite{weedbrook}, the vector of the field
quadratures can be defined as $\hat R = (\hat x^A,\hat p^A, \hat
x^B,\hat p^B)^{\sf T}$, which satisfies the canonical commutation
relations $[{{{\hat R}_k},{{\hat R}_l}} ] = i{\Omega _{kl}}$ with
the symplectic form $\Omega = {{\ 0\ \ 1}\choose{-1\
0}}^{\bigoplus{2}}$. Note that all of the Gaussian properties can be
determined from the symplectic form of the covariance matrix (CM)
$\mathbf{\sigma}_{ij} = \text{Tr}\big[ {{{\{ {{{\hat R}_i},{{\hat
R}_j}} \}}_ + }\ {\rho _{AB}}} \big]$ \cite{RSP,RSP1,RSP2,RSP3},
which is transformed in a standard form through diagonal subblocks
\begin{eqnarray}\label{wsm11}
 \mathbf{{\sigma}}_{AB}= \left(\!\!\begin{array}{cc}
A&C\\
C^T&B
\end{array}\!\!\right) ,
\end{eqnarray}
with $A=a\mathbb{I}$, $B=b\mathbb{I}$, and $C=diag \{c_1,c_2\}$. The
symplectic eigenvalues of $\sigma_{AB}$ are given by
$2{\nu}_{\mp}^2=\Delta \mp\sqrt{\Delta^2-4\det \sigma_{AB}}$, with
$\Delta=\det A+\det B+2\det C$. See
\cite{RSP2,RSP3,gaussian1,gaussian2,gaussian3} for more details
about the structural and formal description of Gaussian quantum
states in phase space.

Now we will briefly introduce the definition of mutual information
(or the total correlation), classical correlation, and quantum
correlation (or the quantum discord) in the continuous variable case
\cite{discord2,discord1,discord3}.

$\textit{Total correlation or mutual information.}$ The mutual
information, or a measurement of the total correlations in a quantum
state, is often used to describe the amount of information in a
quantum state. In general, in this analysis we turn to R\'{e}nyi
entropy of order 2 (instead of the commonly-used von Neumann entropy
in the previous works) to quantify all of the relevant quantities,
which satisfies the strong subadditivity inequality for arbitrary
Gaussian states \cite{inequality} with clear and detailed
expression. Therefore, the total correlation between subsystems $A$
and $B$ can be quantified by the R\'{e}nyi entropy mutual
information
\begin{eqnarray}\label{Total}
\nonumber{I}_2(\sigma_{AB})&=&{S}_2(\sigma_A)+{S}_2(\sigma_B)-{S}_2(\sigma_{AB})\\
&=&\frac{1}{2}\ln\big(\frac{\det A\det B}{\det \sigma_{AB}}\big)\ ,
\end{eqnarray}
with the R\'{e}nyi entropy
${S}_2(\sigma)=\frac{1}{2}\ln(\det \sigma)$ and a given
Gaussian state $\sigma_{AB}$.

$\textit{Classical correlation.}$ It is well known that one-way
classical correlation is always obtained by local measurements.
Specially, when the most informative local measurement is performed
on subsystem $B$, we can define ${J}_2(\sigma_{A|B})$ in
terms of how much the ignorance about the state of subsystem $A$ is
reduced. Based on the newly-introduced Gaussian R\'{e}nyi-2 measure
of one-way classical correlation, ${J}_2(\sigma_{A|B})$
could be interpreted as the maximum decrease in the R\'{e}nyi-2
entropy of subsystem $A$, given a Gaussian measurement (i.e., the
measurement operation mapping Gaussian states into Gaussian state in
the continuous variable system) performed on subsystem $B$ and the
maximization performed over all Gaussian measurements. Now the
expression of ${J}_2(\sigma_{A|B})$ is given by
\begin{eqnarray}\label{Classical}
\nonumber{J}_2(\sigma_{A|B})&=&{S}_2(\sigma_A)-\inf_{\Pi_B(\eta)}{S}_2(\tilde{\sigma}_{A})\\
&=&\sup_{\Pi_B(\eta)} \frac{1}{2}\ln\big(\frac{\det A}{\det \tilde{\sigma}_{A}}\big)\ ,
\end{eqnarray}
where $\Pi_B(\eta)$ denotes the Gaussian measurement on subsystem
$B$. One could obtain CM $\tilde{\sigma}_{A}$ through the
conditional state ($\rho_{A|\eta}$) of subsystem $A$, after a
measurement of $\Pi_B(\eta)$ performed on $B$ with the outcome of
$\eta$. It's worth noting that the Gaussian measurements used in
this analysis specially denote Gaussian positive operator valued
measurements (POVMs), which are executable using linear optics and
homodyne detection \cite{detection}. Swapping the roles of the two
subsystems $A\leftrightarrow B$, one could straightforwardly obtain
the expression of ${J}_2(\sigma_{B|A})$ (see \cite{discord}
for more details).

$\textit{Quantum correlation or quantum discord.}$ The quantum
correlation (or quantum discord), as a measurement of the
quantumness of correlations, is originated from the discrepancy
between two classically equivalent definitions of mutual
information. It is defined as the total minus classical correlations
based on R\'{e}nyi-2 entropy
\begin{eqnarray}\label{quantum}
\nonumber{D}_2(\sigma_{A|B})&=&{I}_2(\sigma_{AB})-{J}_2(\sigma_{A|B})\\
&=&\inf_{\Pi_B} \frac{1}{2}\ln\big(\frac{\det B\det \tilde{\sigma}_{A}}{\det \sigma_{AB}}\big)\ ,
\end{eqnarray}
the calculation result of which can be obtained by swapping the
roles of the two subsystems $A\leftrightarrow B$.

As a final remark, one can clearly see that the total correlation
(${S}_2(\sigma)=\frac{1}{2}\ln(\det \sigma)$) is the sum of
classical correlation (${J}_2(\sigma_{B|A})$) and quantum
correlation ($A\leftrightarrow B$), based on the definitions of
three types of correlations shown above.

\bibliography{sample}

\section*{Acknowledgements (not compulsory)}

This work was supported by National Key R\&D Program of China No.
2017YFA0402600, the National Natural Science Foundation of China
under Grants Nos. 11690023, 11373014, and 11633001, Beijing Talents
Fund of Organization Department of Beijing Municipal Committee of
the CPC, the Strategic Priority Research Program of the Chinese
Academy of Sciences, Grant No. XDB23000000, the Interdiscipline
Research Funds of Beijing Normal University, and the Opening Project
of Key Laboratory of Computational Astrophysics, National
Astronomical Observatories, Chinese Academy of Sciences.

\section*{Author contributions statement}

T.-H. Liu contributed in proposing the idea and original paper
writing. S.-M. Wu contributed in calculating results and comparing
the work with relevant literature. S. Cao contributed in discussing
the new ideas, improving the quality of the paper and organizing the
research.

\section*{Additional information}

Competing financial interests: The authors declare no competing financial interests.

\end{document}